\documentclass[
aip,
onecolumn,
notitlepage,
superscriptaddress,
12pt]{revtex4-2}

\usepackage{graphicx}
\usepackage{amsmath, amssymb, amsthm, amsfonts, bbm, bm, enumitem}

\usepackage{hyperref}
\hypersetup{
    colorlinks,
    linkcolor={black},
    citecolor={blue!80!black},
    urlcolor={blue!80!black}
}

\usepackage[dvipsnames]{xcolor}

\usepackage{braket}
\usepackage{xfrac}
\usepackage{float,caption,subcaption}
\usepackage[justification=centering, font=footnotesize, labelfont=bf]{caption}

\usepackage{hyperref}
\usepackage{xcolor}
\hypersetup{
    colorlinks,
    linkcolor={blue!50!black},
    citecolor={blue!50!black},
    urlcolor={blue!80!black}
}

\theoremstyle{plain}
\newtheorem*{theorem}{Theorem}

\newcommand{\com}[2]{\left[#1,#2\right]}
\newcommand{\e}{{\mathrm{e}}}
\renewcommand{\d}{\mathrm{d}}

\newcommand{\norm}[1]{\vert\vert #1 \vert\vert}
\newcommand{\SI}{appendix}

\begin{document} %%%%%%%%%%%%%%%%%%%%%%%%%%%%%%%%%%%%%%%%%%%%%%%%%%%%%%%%%%%%%%%%

\title{Kapitza's Pendulum as a Classical Prelude to Floquet-Magnus Theory}

\author{Johannes K. Krondorfer}
\email{johannes.krondorfer@gmail.com}
\affiliation{Institute of Experimental Physics, Graz University of Technology, Petersgasse 16, 8010 Graz, Austria}

\author{Maria Kainz}
\affiliation{Institute of Theoretical and Computational Physics, Graz University of Technology, Petersgasse 16, 8010 Graz, Austria}

\author{Matthias Diez}
\affiliation{Institute of Experimental Physics, Graz University of Technology, Petersgasse 16, 8010 Graz, Austria}
\affiliation{Institute of Physics, University of Graz, Universit{\"a}tsplatz 5, 8010 Graz, Austria}

\author{Andreas W. Hauser}
\email{andreas.w.hauser@gmail.com}
\affiliation{Institute of Experimental Physics, Graz University of Technology, Petersgasse 16, 8010 Graz, Austria}

\begin{abstract}
We present a pedagogical introduction to Floquet-Magnus theory through the classical example of Kapitza's pendulum \--- a simple system exhibiting nontrivial dynamical stabilization under rapid periodic driving. By deriving the equations of motion and analyzing the system using Floquet theory and the Magnus expansion, we obtain analytical stability conditions and effective evolution equations. While grounded in classical mechanics, the techniques are directly applicable to periodically driven quantum systems as well. The approach is fully analytical, using only tools from theoretical mechanics, linear algebra, and ordinary differential equations, and is suitable for instruction at the advanced undergraduate or graduate level.
\end{abstract}


\maketitle

\section{Introduction}
Periodically driven systems lie at the heart of many modern developments in physics, ranging from quantum control and optical lattices to driven condensed matter and classical wave systems. In quantum and optical settings, periodic driving can be used to engineer effective couplings, synthetic gauge fields, band structures, and topological phases that have no direct static analog.\cite{Goldman2014_eff,Eckardt2017,Rudner2020} In quantum control, fast modulations can suppress decoherence or implement high-fidelity operations.\cite{Biercuk2009} Classical systems, from mechanical metamaterials to driven plasmas, likewise exhibit new collective behavior under time-periodic modulation.\cite{Trainiti2019,higashikawa2018floquetengineeringclassicalsystems} A particularly prominent example is the Paul trap, where rapidly oscillating electric fields confine charged particles, a principle that also underlies modern trapped-ion quantum technologies.\cite{Paul1990,Haffner2008}

The broad relevance of periodic driving raises a common theoretical question: how can one describe a system whose equations of motion depend explicitly on time while still retaining some of the analytical structure familiar from time-independent problems? For general time-dependent systems this is difficult and often requires numerical methods. For periodically driven systems, however, the repeated time structure allows for a systematic treatment using Floquet theory,\cite{floquet1883,Bukov2015} which separates the dynamics into an effective long-time evolution and a periodic micromotion. The Magnus expansion then provides a perturbative route to compute this effective evolution and thereby regain analytical tractability.\cite{Magnus1954,Blanes2009} Together, these methods turn the time-periodic problem into a controlled effective description while preserving the essential physics of the drive.

\begin{figure}[!t]
    \centering
    \begin{minipage}[b]{0.45\textwidth}
        \centering
        \includegraphics[width=\textwidth]{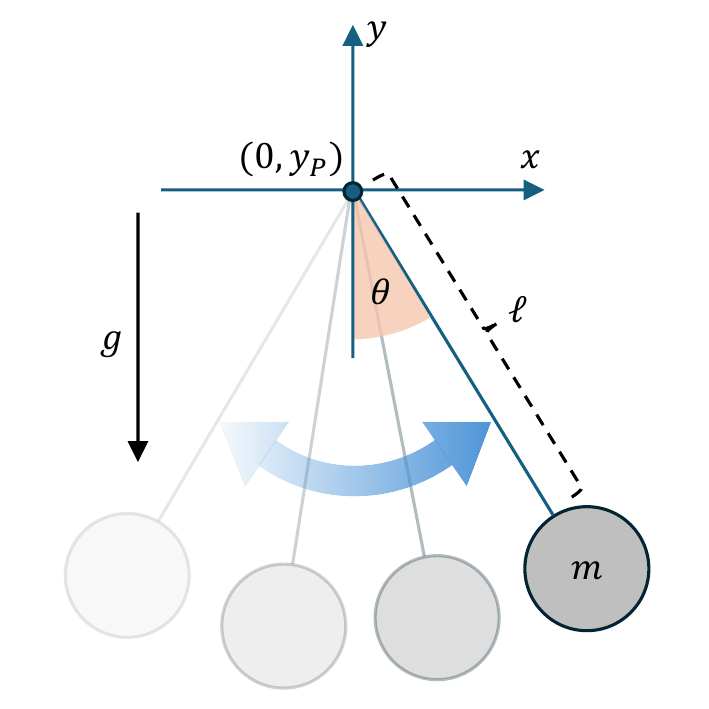}
    \end{minipage}%
    \begin{minipage}[b]{0.45\textwidth}
        \centering
        \includegraphics[width=\textwidth]{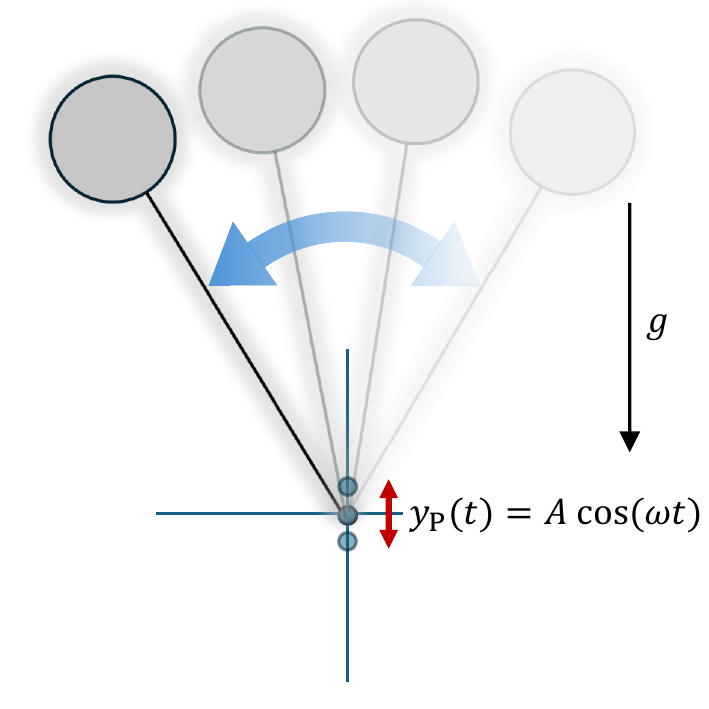}
    \end{minipage}
    \caption{Definition of the coordinates for Kapitza's pendulum in the left panel and illustration of dynamical stabilization for a fast oscillating pivot point in the right panel.}\label{fig:pendulum}
\end{figure}

In this article, we explore Kapitza's pendulum~\cite{Kapitza1951,Butikov2001} \--- a simple pendulum in a homogeneous gravitational field, with a pivot point that oscillates vertically according to $(0,y_\mathrm{P}(t))=(0,A\cos(\omega t))$, as illustrated in Figure~\ref{fig:pendulum}. The Kapitza pendulum is an ideal testbed for exploring time-periodic stability. It combines accessible classical mechanics with rich dynamical behavior and leads to the same mathematical structures that underpin more abstract periodically driven systems~\cite{Bukov2015}.

While this system has long been recognized for its pedagogical value, previous treatments have relied on intuitive reasoning, simplified models of the driving motion, or numerical simulations to illustrate the phenomenon of dynamical stabilization~\cite{kapitza_Pippard_1987, Butikov2001, kapitza_Pal_2023, Butikov_2011}. Our approach follows a systematic and fully analytical route, applying Floquet theory and the Magnus expansion to derive the effective dynamics in a rigorous yet accessible manner, offering a concrete introduction to these techniques.

As shown in the right panel of Figure~\ref{fig:pendulum}, when exposed to periodic vertical oscillations of its pivot point, the Kapitza pendulum can be permanently kept in an upright position -- seemingly defying the law of gravity -- while still performing oscillations for certain ranges of parameter space. The presentation is designed to support instruction at the advanced undergraduate and graduate level, where Kapitza's pendulum can serve as a concrete bridge from familiar classical mechanics to modern theoretical techniques. All derivations are kept analytical, relying only on standard tools from mechanics, linear algebra, and ordinary differential equations. In doing so, we provide a hands-on, conceptually clear route into Floquet-Magnus theory.

The structure of this manuscript is as follows. In Section~\ref{sec:EOM}, we derive the equations of motion for the Kapitza pendulum and linearize them around the stationary points. In Section~\ref{sec:Floquet}, we discuss Floquet theory and provide a simple proof of Floquet's theorem. Section~\ref{sec:Magnus} presents the Magnus expansion and shows how it can be used to approximate the effective dynamics. In Section~\ref{sec:Stability}, we apply these tools to analyze the stability of Kapitza's pendulum under fast periodic driving. Additional derivations and details of the calculations are provided in the \SI.

\section{Lagrange Function and equation of motion}\label{sec:EOM}
We begin by deriving the equation of motion for the Kapitza pendulum with a vertically oscillating pivot point, $y_\mathrm{P}(t) = A\cos(\omega t)$. We define $\theta$ as the angle measured from the downward vertical direction, i.e. from the negative $y$-axis, to the pendulum rod. The pendulum bob has mass $m$ and is attached to a rigid, massless rod of length $\ell$, as illustrated in the left panel of Figure~\ref{fig:pendulum}.

To express the dynamics in terms of the angle $\theta$, we compute the kinetic and potential energy. The coordinates of the pendulum bob are
\begin{equation}
x = \ell \sin(\theta), \quad y = y_\mathrm{P}(t) - \ell \cos(\theta),
\end{equation}
so the potential energy becomes
\begin{equation}
    V = mg\left(y_\mathrm{P} - \ell \cos(\theta) \right)\,.
\end{equation}
The kinetic energy is given by
\begin{equation}
    T = \frac{m}{2}(\dot{x}^2 + \dot{y}^2) = \frac{m\ell^2}{2} \dot{\theta}^2 + m\ell \dot{y}_\mathrm{P} \sin(\theta) \dot{\theta} + \frac{m}{2} \dot{y}_\mathrm{P}^2\,.
\end{equation}
The total Lagrangian is $L = T - V$, yielding
\begin{equation}
    L = \frac{m\ell^2}{2} \dot{\theta}^2 + m\ell \dot{y}_\mathrm{P}(t) \sin(\theta)\, \dot{\theta} + mg\ell \cos(\theta) + \left( \frac{m}{2} \dot y_\mathrm{P}(t)^2 - mg\, y_\mathrm{P}(t) \right)\,,
\end{equation}
where we can neglect the term in brackets, as it does not depend on $\theta$ or $\dot{\theta}$, and therefore does not affect the equation of motion.

With that, we can derive the equation of motion for $\theta$ by utilizing the Euler-Lagrange equation. This yields
\begin{align}
\begin{split}
    0 = \frac{\d}{\d t}\frac{\partial L}{\partial\dot{\theta}} - \frac{\partial L}{\partial\theta} = m\ell^2\ddot{\theta} + m\ell ( \ddot{y}_\mathrm{P} + g ) \sin(\theta)\,,
\end{split}
\end{align}
which simplifies to
\begin{equation}\label{eq:full ode}
    \ddot{\theta} = -\frac{\ddot{y}_\mathrm{P} + g}{\ell}\sin(\theta)\,.
\end{equation}
For completeness, we remark that the same equation of motion can also be obtained by applying Newton’s law in the non-inertial frame attached to the oscillating pivot, where the fictitious forces give rise to Equation~\ref{eq:full ode} directly.
This shows that the vertical acceleration of the pivot, $\ddot{y}_\mathrm{P}(t)$, effectively modifies the gravitational acceleration, and thus alters the stability of the system. In particular, a rapidly oscillating pivot can dynamically stabilize the otherwise unstable inverted position, as we will see below.

\subsection{Stationary points for a constant pivot point}

As a warm-up, we first analyze the undriven case, where the pivot remains fixed at $y_\mathrm{P} = \text{const}$ and thus $\ddot{y}_\mathrm{P} = 0$. The equation of motion simplifies to
\begin{equation}
    \ddot{\theta} = -\frac{g}{\ell} \sin(\theta)\,,
\end{equation}
which describes a simple pendulum in a uniform gravitational field. Stationary (equilibrium) points occur when the angular acceleration and the angular velocity vanish, i.e. $\ddot{\theta} = \dot{\theta} = 0$, giving
\begin{equation}
    \sin(\theta) = 0 \quad \Rightarrow \quad \theta_\mathrm{s} = n\pi, \qquad n \in \mathbb{Z}\,.
\end{equation}
Thus, the pendulum has two types of equilibrium: the downward vertical position $\theta_\mathrm{s} = 0 \mod 2\pi$, and the inverted vertical position $\theta_\mathrm{s} = \pi \mod 2\pi$.

To assess the stability of these stationary points, we linearize the equation of motion around each point by using a first order Taylor approximation of $\sin(\theta)$. We write $\theta(t) = \theta_\mathrm{s} +  \delta(t)$ with $\delta(t) \ll 1$ and obtain
\begin{align}
    \sin(\theta_\mathrm{s} + \delta) &= \sin(\delta) \approx \delta
\end{align}
for $\theta_\mathrm{s} = 0$ and 
\begin{align}
    \sin(\theta_\mathrm{s} + \delta) &= \sin(\pi + \delta) \approx -\delta
\end{align}
for $\theta_\mathrm{s} = \pi$.

\begin{itemize}
    \item Expansion around $\theta_\mathrm{s} = 0$: For the downward equilibrium, the linearized equation of motion becomes $\ddot{\delta} = -\frac{g}{\ell} \delta$ which describes a harmonic oscillator with natural frequency 
    \begin{equation}\label{eq:omega 0}
        \omega_0 = \sqrt{\frac{g}{\ell}}\,.
    \end{equation}
    This solution is oscillatory and bounded, implying that $\theta_\mathrm{s} = 0$ is a stable equilibrium.

    \item Expansion around $\theta_\mathrm{s} = \pi$: For the inverted equilibrium, the linearized equation of motion becomes $\ddot{\delta} = \frac{g}{\ell} \delta$ which describes an inverted harmonic oscillator. The solutions are exponential functions that grow in time, indicating that $\theta_\mathrm{s} = \pi$ is an unstable equilibrium.
\end{itemize}
This result reflects our intuition: when the pendulum hangs downward, small displacements lead to restoring forces that return it to equilibrium. When it points upward, any small deviation leads to an increasing departure from the vertical position. These features are captured by the sign of the linearized force term. This classical behavior sets the stage for the more intriguing case of an oscillating pivot, where the inverted position may become dynamically stabilized \--- a striking departure from static intuition. In the next subsection, we extend this analysis to incorporate time-periodic forcing and analyze how stability changes in the remainder of this manuscript.

\subsection{Linearization of Kapitza's Pendulum}\label{sec:lin kapitza}
We now return to the full time-dependent case, where the pivot point oscillates vertically as $y_\mathrm{P}(t) = A \cos(\omega t)$, as stated in Equation~\ref{eq:full ode}.
Despite the explicit time dependence, the stationary points of the system remain the same as in the undriven case: $\theta_\mathrm{s} = 0\mod 2\pi$ and $\theta_\mathrm{s} = \pi \mod 2\pi$. This is because the right-hand side of the equation of motion vanishes whenever $\sin(\theta) = 0$, independently of the driving $\ddot{y}_\mathrm{P}(t)$. However, the stability of these points can no longer be deduced directly from the sign of a constant coefficient, as in the static case. Instead, it must be assessed more carefully due to the time-dependent nature of the system.

To analyze the local dynamics near each equilibrium, we linearize the equation of motion for small angular deviations by writing $\theta(t) = \theta_\mathrm{s} +  \delta(t)$ with $\delta(t) \ll 1$ as before. Substituting this into the equation of motion and using the first order Taylor approximation around each equilibrium, we obtain the linearized differential equation
\begin{equation}\label{eq:Kapitza HO}
    \ddot{\delta}(t) = \mp \frac{\ddot{y}_\mathrm{P}(t) + g}{\ell}\, \delta(t)\,,
\end{equation}
where the $+$ sign corresponds to the inverted position $\theta_\mathrm{s} = \pi$, and the $-$ sign to the downward equilibrium $\theta_\mathrm{s} = 0$.

This equation describes a driven harmonic oscillator with time-periodic coefficients. The mass has dropped out of the equation of motion. The pendulum length $\ell$ and natural frequency $\omega_0 = \sqrt{g/\ell}$ provide the relevant scales for the driving function $\ddot y_\mathrm{P}(t)$, which determines the stability properties. The standard route to solving this problem --- which is not the one that will be followed in this paper --- is therefore to transform Equation~\ref{eq:Kapitza HO} into dimensionless variables by choosing $\tau = \omega t/2$, which leads to the well-known Mathieu equation~\cite{Yakubovich1975, Butikov2001, Butikov_2011}
\begin{equation}\label{eq:mathieu}
    \frac{\d^2}{\d\tau^2} \delta(\tau) = -(a - 2q\cos(2\tau))\delta(\tau) \quad\text{with}\quad a = \pm \frac{4\omega_0^2}{\omega^2} \quad\text{and} \quad q=\pm\frac{2A}{\ell}\,.
\end{equation}
The usual Mathieu-equation treatment then analyzes stability perturbatively in $q$, often together with an averaging over the fast oscillations.\cite{Yakubovich1975} For the full nonlinear pendulum, a related heuristic approach separates the slow and fast components of the motion and derives an averaged effective potential for the stabilization mechanism.\cite{Kapitza1951, Butikov2001} Here, instead, we use Floquet theory and the Magnus expansion to obtain the effective evolution systematically.

\section{Floquet Theory}\label{sec:Floquet}
To study the behavior of Kapitza's pendulum within Floquet-Magnus theory, it is convenient to transform the second-order ODE in Equation~\ref{eq:Kapitza HO} into a system of first-order equations
\begin{equation}
    \frac{\mathrm{d}}{\mathrm{d}t} 
    \begin{bmatrix}
        \delta \\ \dot{\delta}
    \end{bmatrix} = 
    \begin{bmatrix}
        0 & 1 \\
        \alpha_\mp(t) & 0
    \end{bmatrix}
    \begin{bmatrix}
        \delta \\ \dot{\delta}
    \end{bmatrix}, \qquad \text{with }\quad \alpha_\mp(t) := \mp \frac{\ddot{y}_\mathrm{P} + g}{\ell}\,,
\end{equation}
or more compactly, $\frac{\d}{\d t} \bm{\delta}(t) = H(t) \bm{\delta}(t) $, where $H(t)$ is the time-dependent coefficient matrix of the first-order system. For a given initial condition $\bm{\delta}(t_0) = \bm{\delta}_0$ the solution is unique when the coefficient matrix is sufficiently smooth. It is useful to separate this initial vector from the dynamics itself. For a linear system, this can be done by introducing a matrix-valued function $U(t,t_0)$, called the propagator, that maps the initial vector at time $t_0$ to the solution at time $t$,
\begin{equation}
    \bm{\delta}(t) = U(t,t_0)\bm{\delta}_0\,.
\end{equation}
Once $U(t,t_0)$ is known, the solution for any initial condition follows by matrix-vector multiplication. Substituting this form into $\frac{\d}{\d t} \bm{\delta}(t) = H(t) \bm{\delta}(t)$ and using that $\bm{\delta}_0$ can be arbitrary gives the matrix differential equation
\begin{equation}\label{eq:propagator}
    \frac{\d}{\d t} U(t,t_0) = H(t)\, U(t,t_0), \qquad U(t_0,t_0) = I,
\end{equation}
with identity matrix $I$. The initial condition $U(t_0,t_0)=I$ expresses that the state is unchanged at the initial time. Thus the propagator isolates the dynamics governed by $H(t)$ from the choice of initial condition.

For Kapitza's pendulum, the coefficient matrix $H(t)$ is time-periodic because the pivot motion is periodic. We can therefore apply Floquet theory to analyze its stability.\cite{floquet1883, Chicone2024Floquet, Bukov2015} In this setting, the one-period evolution operator $U(t_0+T,t_0)$, also called the monodromy matrix, plays a central role as we will see below. Floquet theory shows that the propagator can be decomposed into two parts: a purely exponential evolution governed by an effective time-independent matrix, and a time-periodic modulation. This standard result for linear differential equations with periodic coefficients is formalized in Floquet's theorem.\cite{Chicone2024Floquet} To keep the discussion self-contained, we provide a short proof below that uses only standard ideas from ordinary differential equations and linear algebra.

\begin{theorem}[Floquet's Theorem]
    Consider a linear homogeneous differential equation with coefficient matrix $H$, as in Equation~\ref{eq:propagator}, where $H$ is a continuous, $T$-periodic matrix-valued function, i.e.\ $H(t+T) = H(t)$ for all $t$. Then the propagator $U(t,t_0)$ admits the decomposition
    \begin{equation}\label{eq:floquet}
        U(t,t_0) = P(t,t_0)\, e^{\widetilde{H}(t_0)(t-t_0)},
    \end{equation}
    where $P(t,t_0)$ is a matrix-valued function that is $T$-periodic in its first argument, and $\widetilde{H}(t_0)$ is a constant matrix known as the Floquet Hamiltonian with initial time $t_0$, given by
    \begin{equation}
        \widetilde{H}(t_0) = \frac{1}{T} \log U(t_0+T,t_0).
    \end{equation}
\end{theorem}

\subsection{Proof of Floquet's Theorem}
We prove the theorem for $t_0 = 0$ and omit the initial time as an explicit argument. The general case can be established analogously. A brief discussion on the influence of the initial time is provided in the \SI.

With $\widetilde{H} = T^{-1}\log U(T)$, we define $P(t) := U(t)\, \e^{-\widetilde{H}t}$. It remains to show that $P(t)$ is $T$-periodic.

We first establish the identity $U(t+T)=U(t)U(T)$. To see this, define $V(t) := U(t+T) U^{-1}(T)$. Because $U^{-1}(T)$ is independent of $t$ and $H(t)$ is $T$-periodic, $V(t)$ satisfies the same differential equation as $U(t)$
\begin{equation}
    \frac{\d}{\d t} V(t) = \frac{\d}{\d t} U(t+T) U^{-1}(T) = H(t+T) U(t+T) U^{-1}(T) = H(t) V(t)\,,
\end{equation}
with the same initial condition, $V(0) = U(T) U^{-1}(T) = I=U(0)$. By uniqueness of solutions, $V(t) \equiv U(t)$, and therefore $U(t+T) = U(t) U(T)$.

We can now compute
\begin{align}
\begin{split}
    P(t+T) 
    &= U(t+T)\, \e^{-\widetilde{H}(t+T)} = U(t) U(T) \,\e^{-\widetilde{H}T}\, \e^{-\widetilde{H}t} \\
    &= U(t) U(T) U^{-1}(T)\, \e^{-\widetilde{H}t} = U(t)\, \e^{-\widetilde{H}t} = P(t)\,,
\end{split}
\end{align}
where we used $\e^{-\widetilde{H}T} = \e^{-\log U(T)} = U^{-1}(T)$. Thus $P(t)$ is $T$-periodic, and the definition of $P(t)$ immediately gives $U(t) = P(t) \, \e^{\widetilde{H}t}$, which proves the theorem.

This decomposition separates the evolution into a global exponential part $\e^{\widetilde{H}t}$ and a local time-periodic modulation $P(t)$. While $P(t)$ captures the short-time periodic motion within one driving period, often called micromotion, $\widetilde{H}$ governs the long-term dynamics. In the time-independent case, this reduces to the usual matrix-exponential solution, since $P(t)=I$, $\widetilde{H}=H$, and $U(t)=\e^{Ht}$.

\subsection{The Floquet Hamiltonian and Stability}\label{sec:floquet stability}
To understand how the decomposition $U(t) = P(t)\, \e^{\widetilde{H}t}$ helps to analyze the stability of $U$, one can look into the transformed variable $W(t):= P(t)^{-1} U(t) = \e^{\widetilde{H}t}$. As immediately seen from the definition, the transformed variable $W$ obeys the evolution equation
\begin{equation}
    \frac{\d}{\d t} W(t) = \widetilde{H}\, W(t)\,,
\end{equation}
where the evolution matrix is given by the Floquet Hamiltonian $\widetilde{H}$. Since $W$ is simply the matrix exponential of $\widetilde{H}$, $W$ is stable if and only if the spectrum (the set of eigenvalues) of $\widetilde{H}$ is a subset of the left half plane, i.e.
\begin{equation}
    \sigma(\widetilde{H})\subseteq \{z\in\mathbb{C}\,\vert\, \mathrm{Re}(z) < 0 \}.
\end{equation}
To relate this property of $W$ to the stability of $U$ we note that $P$ is periodic, invertible, and continuous, and therefore, one can make estimates on the norm of $U$
\begin{align}
    \norm{U(t)} &= \norm{P(t)W(t)} \leq \norm{P(t)}\, \norm{W(t)} \leq \max_{\tau\in[0,T]}\norm{P(\tau)}\, \norm{W(t)}
\end{align}
\begin{align}
\begin{split}
    \norm{W(t)} &= \norm{P(t)^{-1} U(t)} \leq \norm{P(t)^{-1}}\, \norm{U(t)} \\
    &\Rightarrow \norm{U(t)} \geq \norm{P(t)^{-1}}^{-1}\, \norm{W(t)} \geq \min_{\tau\in[0,T]} \norm{P(\tau)^{-1}}^{-1}\, \norm{W(t)}\,,
\end{split}
\end{align}
which yields
\begin{equation}
    C'\, \norm{W} \leq \norm{U} \leq C''\, \norm{W}\,,
\end{equation}
for some constants $C'$ and $C''$. This means that the stability and long-term behavior of $U$ is determined by $W$ and thus by $\widetilde{H}$, implying that $U$ is bounded (i.e., the system is stable) if and only if $W$ is bounded.

\subsection{Physical Interpretation}
In the context of the Kapitza pendulum, the Floquet Hamiltonian $\widetilde{H}$ captures the effective dynamics of the system under rapid periodic driving. While $H(t)$ contains the full time-dependent behavior, $\widetilde{H}$ offers a simplified description analogous to an averaged or ''dressed'' evolution. This interpretation becomes especially powerful in the high-frequency regime with small amplitude, where the fast driving causes $P(t)$ to oscillate rapidly around the identity. In this limit, $\widetilde{H}$ approximates the net effect of the driving on slower time scales and serves as the cornerstone for the concept of dynamical stabilization. 

But we still need to compute $\widetilde{H}$ explicitly. Therefore, in the next section, we discuss the Magnus expansion, a method for systematically computing $\widetilde{H}$ via perturbation theory in the driving strength and inverse frequency.

\section{Dyson Series and Magnus Expansion}\label{sec:Magnus}
Having established the central role of the one-period evolution operator $U(T)$ and the associated Floquet Hamiltonian $\widetilde{H}$, we now turn to their computation. In most cases, however, analytical expressions for these quantities are out of reach, necessitating either numerical methods or suitable approximations. To deepen the conceptual understanding of periodically driven systems, it is desirable to develop analytical approximations. In the following, we therefore explore systematic methods to approximate $\widetilde{H}$ and $U(T)$.

\subsection{The Dyson Series}
One common approach is the Dyson series~\cite{Dyson1949}, also known in ordinary differential equation theory as Picard iteration~\cite{Coddington1955}. It is widely used in time-dependent perturbation theory, especially for time-dependent quantum systems, and expands the propagator $U(t)$ in powers of the coefficient matrix $H(t)$
\begin{equation}\label{eq:dyson}
    U(t) = \sum_{k=0}^\infty U^{(k)}(t)\,, \qquad U^{(0)}(t) = I\,,
\end{equation}
where $I$ is the identity matrix.
Each term is given by a nested time integral
\begin{equation}\label{eq:dyson-k}
    U^{(k)}(t) = \int_0^t \mathrm{d}t_1 \int_0^{t_1} \mathrm{d}t_2 \cdots \int_0^{t_{k-1}} \mathrm{d}t_k\, H(t_1) H(t_2) \cdots H(t_k)\,.
\end{equation}
For coefficient matrices commuting at all different times, i.e. $\com{H(t)}{H(t')}=0$ for all $t, t'$, the Dyson series reduces to the usual matrix exponential (see the \SI). For the Kapitza problem considered here, however, the coefficient matrices $H(t)$ and $H(t')$ do not commute in general, so this simplification is not available.

While the Dyson series offers a formally correct perturbative expansion, truncating it at finite order can spoil important structural properties of the exact time evolution. In quantum systems, for example, a truncated Dyson series generally breaks unitarity, so norms and probabilities are not preserved exactly. In classical Hamiltonian systems, finite-order truncation can break symplecticity, and thus the Hamiltonian phase-space structure.\cite{Blanes2009} This motivates an approximation scheme that keeps the evolution in exponential form.

\subsection{The Magnus Expansion}
The Magnus expansion provides such a representation,
\begin{equation}
    U(t) = \exp\left( \Omega(t) \right)\,,
\end{equation}
where the exponent $\Omega(t)$ is itself expanded as a series
\begin{equation} \label{eq:omega}
    \Omega(t) = \sum_{k=1}^{\infty} \Omega^{(k)}(t), \qquad \Omega^{(k)} = \mathcal{O}(\|H\|^k).
\end{equation}
The exact Magnus series is generally infinite, so practical calculations still require truncation. Its advantage over a truncated Dyson series is that a finite-order Magnus approximation remains a proper exponential and therefore preserves structural properties more faithfully.\cite{Magnus1954,Blanes2009}

As we will see in Section~\ref{sec:kapitza magnus derivation}, the third-order expansion gives the first nontrivial correction for the Kapitza pendulum. For this reason, we compute the Magnus terms only up to third order here, while noting that any order can be obtained with the approach outlined below.

Although closed-form expressions for each $\Omega^{(k)}$ exist in terms of iterated commutators and integrals (see Refs.~\citenum{Blanes2009, Magnus1954}), it is simpler for explicit hand calculations to compute the Magnus terms by matching the Magnus series to the Dyson expansion order by order. Starting from $U(t)=\exp(\Omega(t))$ and substituting Equations~\ref{eq:dyson} and~\ref{eq:omega}, we write
\begin{equation}\label{eq:magnus-dyson-short}
    \exp\left(\sum_{k=1}^{\infty}\Omega^{(k)}(t)\right)
    = \sum_{k=0}^{\infty} U^{(k)}(t)\,.
\end{equation}
By expanding the exponential and matching terms of equal order in $H$, the Magnus terms can be expressed through Dyson terms. The detailed calculation is given in the \SI; the result up to third order reads
\begin{align}\label{eq:magnus 3}
\begin{split}
    \Omega^{(1)}(t) = U^{(1)}(t)\,, \\
    \Omega^{(2)}(t) = U^{(2)}(t) - \frac{1}{2} (\Omega^{(1)}(t))^2 \,, \\
    \Omega^{(3)}(t) = U^{(3)}(t) - \frac{1}{2}\left( \Omega^{(1)}(t) \Omega^{(2)}(t) + \Omega^{(2)}(t) \Omega^{(1)}(t)  \right) - \frac{1}{6} (\Omega^{(1)}(t))^3\,.
\end{split}
\end{align}

In the context of Floquet theory, we are particularly interested in computing the one-period evolution $U(T)$ and extracting the effective generator
\begin{equation}
    \widetilde{H} = \frac{1}{T} \Omega(T),
\end{equation}
as we will do in the next section to identify the conditions for dynamical stabilization for the Kapitza pendulum.

\section{Stability Analysis of Kapitza's Pendulum}\label{sec:Stability}
\subsection{Analytical Derivation of the Effective Evolution Equation}\label{sec:kapitza magnus derivation}
We are now in a position to analyze the stability of Kapitza's pendulum using the tools introduced in the previous sections. In the linearized regime, the dynamics near the fixed points $\theta_\mathrm{s} = 0, \pi$ are governed by the time-dependent coefficient matrix
\begin{equation}
    H(t) = 
    \begin{bmatrix}
        0 & 1 \\
        \alpha_\mp(t) & 0
    \end{bmatrix}, \qquad \text{with} \quad \alpha_\mp(t) = \mp \frac{\ddot{y}_\mathrm{P}(t) + g}{\ell}.
\end{equation}
We specialize to the case where the pivot oscillates vertically as $y_\mathrm{P}(t) = A\cos(\omega t)$ with small amplitude $A$ and high frequency $\omega$, such that $A / \ell \ll 1$ and $\omega_0^2 / \omega^2 \ll 1$, while $A\omega / \ell$ is of the same order as $\omega_0 = \sqrt{g/\ell}$, which will turn out as the region where dynamic stabilization is possible.

As shown in Section~\ref{sec:Floquet}, the long-time behavior of this periodically driven system is governed by the Floquet Hamiltonian $\widetilde{H}$, defined via the one-period time-evolution matrix $U(T)$, with $T = 2\pi / \omega$, as
\begin{equation}
    \widetilde{H} = \frac{1}{T} \log U(T).
\end{equation}
While $U(T)$ cannot be computed in closed form, we approximate $\widetilde{H}$ using the Magnus expansion up to third order
\begin{equation}
    \widetilde{H} \approx \frac{1}{T} \left( \Omega^{(1)}(T) + \Omega^{(2)}(T) + \Omega^{(3)}(T) \right).
\end{equation}
Evaluating the integrals for the Magnus terms (see the \SI), we obtain the leading-order effective Hamiltonian
\begin{equation}
    \widetilde{H} =
    \begin{bmatrix}
        0 & 1 \\
        -\frac{1}{2}\left(\frac{A\omega}{\ell}\right)^2 \mp \omega_0^2 & 0
    \end{bmatrix}\,,
\end{equation}
This matrix governs the long-term evolution of the pendulum.
To analyze stability, we convert the above first-order system back into a second-order equation. Letting $\delta(t)$ denote the angular deviation from $\theta_s = 0$ or $\theta_s = \pi$, the effective dynamics read
\begin{equation}\label{eq:eff model}
    \ddot{\delta} + \left( \frac{1}{2} \left( \frac{A\omega}{\ell} \right)^2 \pm \omega_0^2 \right) \delta = 0\,.
\end{equation}
This is a harmonic oscillator with a square frequency
\begin{equation}
    \omega_\pm^2 = \frac{1}{2} \left( \frac{A\omega}{\ell} \right)^2 \pm \omega_0^2\,,
\end{equation}
and as such the system is stable if and only if $\omega_\pm^2 > 0$. This condition leads to the following conclusions:
\begin{itemize}[leftmargin=*]
    \item For the downward equilibrium $\theta_s = 0$, we have $\omega_+^2 > 0$ for all values of $A$ and $\omega$, so the motion remains stable under fast driving, but with a different oscillation frequency.
    
    \item For the inverted equilibrium $\theta_s = \pi$, the condition $\omega_-^2 > 0$ gives a nontrivial constraint on the driving parameters, $\left( A\omega / \ell \right)^2 > 2 \omega_0^2$, which we can compactly write as
    \begin{equation}\label{eq:kapitza stability}
        -\!q^2 < 2a\,,
    \end{equation}
    with $q = -2A/\ell$ and $a = -4\omega_0^2/\omega^2$, as in Section~\ref{sec:lin kapitza}.
    This gives the stability condition for dynamical stabilization: if the driving is sufficiently fast and strong, the inverted pendulum becomes effectively stable \--- even though it is statically unstable. Physically, this stabilization can be understood in the non-inertial frame where the pivot is at rest: the pendulum bob experiences an oscillating inertial force which averages to zero, but its torque does not, since the force and lever arm oscillate in phase. As discussed clearly by Butikov,\cite{Butikov2001} this non-vanishing average torque produces the effective potential that stabilizes the inverted position.
\end{itemize}

This striking phenomenon, known as dynamical stabilization, is a hallmark of systems governed by effective time-averaged dynamics. It emerges naturally from the Floquet-Magnus framework and can be understood purely from the sign of the effective potential. In the case of the inverted pendulum, the classically unstable upward position is stabilized by rapid periodic driving of the pivot point.

\subsection{Comparison of Effective and Exact Dynamics}
A standard way to illustrate the stability of the Mathieu equation (cf. Equation~\ref{eq:mathieu}), and hence of the Kapitza pendulum, is through the Ince–Strutt diagram\cite{Yakubovich1975}, which depicts stable and unstable regions in the parameter space $(a,q)$. The exact stability regions can be obtained by computing the one-period evolution operator and determining its stability, as discussed in Section~\ref{sec:floquet stability}. Since the sign of $q$ does not change the stability, we only consider positive values. 

\begin{figure}[!t]
    \centering
    \includegraphics[width=\textwidth]{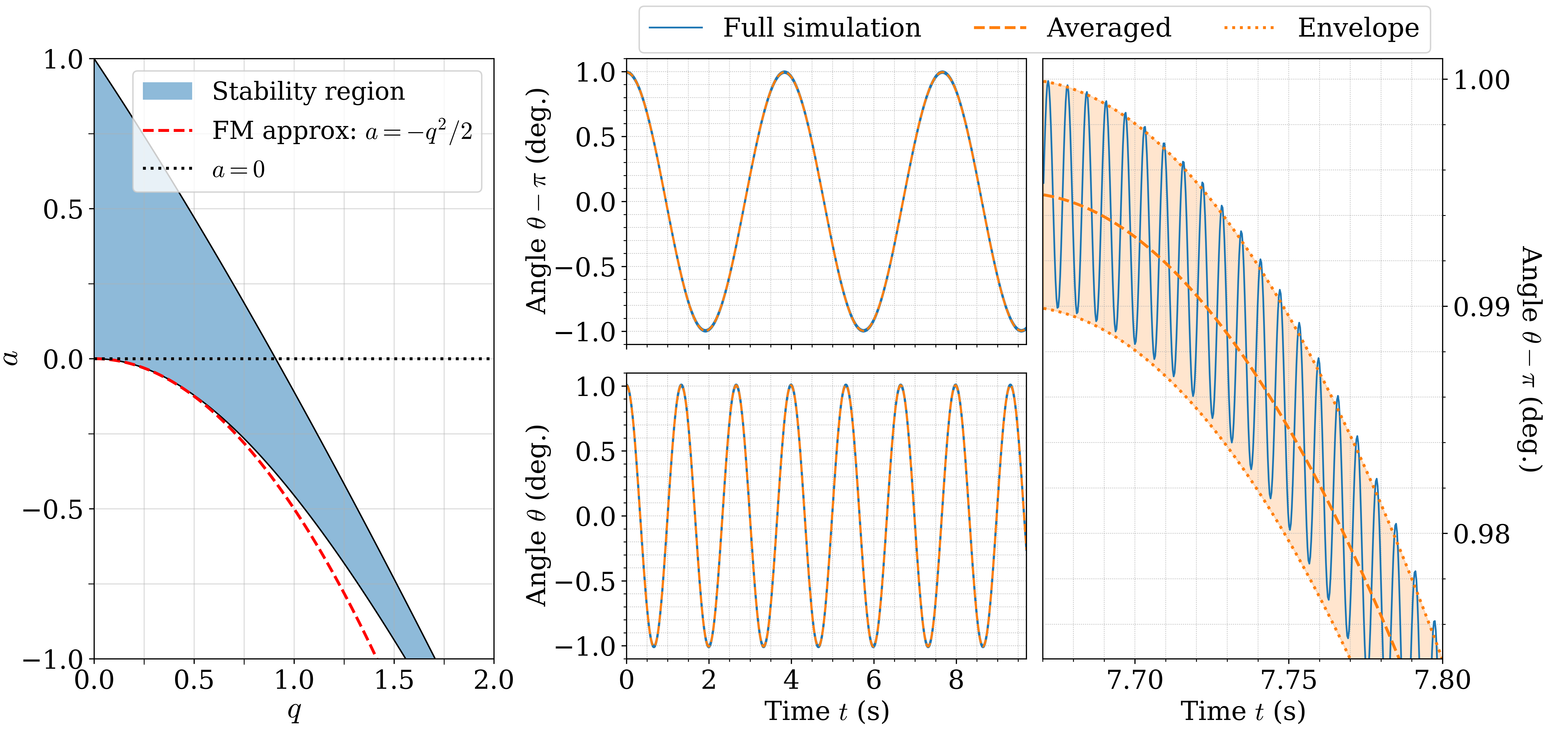}
    \put(-465,210){\small (a)}
    \put(-302,210){\small (b)}
    \vspace{-0.5cm}
    \caption{(a) Ince--Strutt diagram of the linearized Kapitza pendulum near the high-frequency, small-amplitude regime. The dashed line shows the approximate stability boundary from Equation~\ref{eq:kapitza stability}, while the shaded regions mark the exact stability domains obtained from the Floquet analysis of the Mathieu equation. The dotted line marks the \(a=0\) boundary, which separates the downward-pendulum regime \((a>0)\) from the inverted-pendulum regime \((a<0)\).
    (b) Full nonlinear dynamics compared with the effective linearized model near the inverted (upper panel) and downward (lower panel) equilibria for parameters satisfying the high-frequency and small-amplitude condition. The right panel shows a zoomed-in view of the oscillations around the inverted position.}\label{fig:numeric}
\end{figure}

In Figure~\ref{fig:numeric}a we show the corresponding stability diagram in the vicinity of the small-amplitude and high-frequency regime, i.e. $q\ll1$ and $a\ll1$. The shaded regions mark the stability domains obtained from the exact Floquet analysis of the Mathieu equation, while the dashed line shows the approximate stability boundary, derived in Equation~\ref{eq:kapitza stability}. The stability diagram is divided in two regions:
\begin{itemize}
    \item The upper region, $a>0$, corresponds to the downward equilibrium, which is stable for all parameters provided that the high-frequency and small-amplitude conditions are satisfied. When these conditions are violated also the downward equilibrium can become unstable, as we can see for larger values of $a$ and $q$.

    \item The lower region, $a<0$, also called Kapitza region, corresponds to the inverted equilibrium, which is unstable in the absence of driving ($q=0$), but becomes stable when the driving strength and frequency are sufficiently large. The approximate stability boundary from Equation~\ref{eq:kapitza stability} agrees well with the exact stability boundary in the small-amplitude and high-frequency regime, while deviations become more pronounced for larger values of $a$ and $q$.
\end{itemize}

In Figure~\ref{fig:numeric}b, we compare the time evolution of the exact and effective dynamics near the inverted ($\theta \approx \pi$) and downward ($\theta \approx 0$) equilibrium. The full nonlinear dynamics are obtained by numerically solving Equation~\ref{eq:full ode}, while the effective solution is obtained analytically by solving the effective harmonic oscillator system given in Equation~\ref{eq:eff model}. Parameters are chosen to satisfy the high-frequency and small-amplitude conditions under which the analytical approximation is valid ($g=9.81\;\mathrm{m/s^2}$, $\ell = 1\;\mathrm{m}$, $A=5\;\mathrm{mm}$, $\omega = 1000\;\mathrm{s^{-1}}$). The numerical simulations were carried out using standard ODE solvers ($\texttt{solve\_ivp}$ in Python); see the \SI{} for annotated source code. The right panel shows a zoomed-in view of the oscillations around $\theta \approx \pi$. We observe that the effective solution captures the averaged evolution of the fast-oscillating full dynamics and provides an accurate description in the regime of small amplitude, high-frequency driving, and small deviations from the equilibrium, i.e. the domain in which the linearization and high-frequency expansion are valid. The shaded band illustrates the bounded micromotion around the effective evolution. In this case, the upper and lower bounds of the shaded band can also be obtained by changing the initial time in the Floquet decomposition, which corresponds to a gauge transformation of the Floquet Hamiltonian, as discussed in the \SI. Similarly, all intermediate curves inside the band can be represented by a suitable Floquet Hamiltonian for a suitable choice of initial time. A detailed discussion of this Floquet-gauge freedom is beyond the scope of this manuscript; for further details, see Ref.~\citenum{Bukov2015}. Therefore, the averaged solution is a convenient representative of the effective long-time dynamics.

\section{Conclusion and Outlook}
The analysis above shows how the dynamical stabilization of Kapitza's pendulum follows from Floquet theory and the Magnus expansion. The example is useful because the formalism has a direct physical interpretation: rapid periodic driving leads to an effective averaged dynamics, while the micromotion accounts for the remaining fast oscillations. This provides an accessible entry point to the basic language of driven systems.

Moving to quantum dynamics, a closely related example is a driven quantum harmonic oscillator. The oscillator remains familiar, but the time evolution is generated by a time-dependent quantum Hamiltonian. The same ideas can then be discussed in terms of the quantum propagator, effective Hamiltonians, and micromotion.

Similar concepts already appear in simple research models, such as driven quantum two-level systems. A periodic laser or microwave field can couple the two states and generate effective couplings, frequency shifts, and controlled population dynamics. This connects to coherent control and dynamical decoupling.\cite{Viola1998,Biercuk2009} Related periodically modulated optical fields can address level transitions in atoms and molecules,\cite{krondorfer_nuclear_2023,krondorfer_optical_2024} and can be used to control two-level or even multilevel systems.\cite{krondorfer2025opticalnuclearelectricresonance,krondorfer2025singlequditcontrol87sr}

In larger systems, this becomes the idea of Floquet engineering: the drive is chosen to design an effective Hamiltonian.\cite{Goldman2014_eff,Eckardt2017} In optical lattices, laser fields form the lattice potential, and periodic shaking or modulation of these fields can change tunneling rates and create synthetic gauge fields or tunable band structures.\cite{Aidelsburger2011,Eckardt2017} In solid-state systems, periodic electromagnetic fields can reshape electronic bands and generate topological edge states without a static counterpart.\cite{Oka2009,Lindner2011,Rudner2020} Classical driven media, such as elastic and phononic structures, provide analogous examples where time-periodic modulation changes wave propagation and stability.\cite{Trainiti2019,Leonard2018,higashikawa2018floquetengineeringclassicalsystems} Thus, Kapitza's pendulum is a simple entry point into the broader study of driven classical and quantum systems.

\appendix

\section*{Overview of the Appendix}
The \SI{} contains additional details supporting the main text. It includes a discussion of the Floquet gauge freedom associated with the choice of initial time, a short derivation of the Dyson series for commuting matrices, the order-by-order matching of the Dyson and Magnus expansions up to third order, and the detailed calculation of the effective Floquet Hamiltonian for Kapitza's pendulum.

\section{Gauge Freedom in Floquet Theory}\label{app:gauge}
Floquet theory admits a gauge freedom associated with the choice of the initial time $t_0$. This choice fixes the phase of the drive at which the one-period evolution is evaluated, but does not change the physical dynamics or the driving period. Let $\widetilde H(t_0)$ and $\widetilde H(t_1)$ denote the Floquet Hamiltonians obtained for initial time $t_0$ and $t_1$, respectively. Using the group property of the propagator $U(\tau_2, \tau_0) = U(\tau_2,\tau_1) U(\tau_1,\tau_0)$, and the periodicity $U(t_1+T,t_0+T) = U(t_1,t_0)$ we can relate the one-period evolution operator for different initial times via
\begin{align}
\begin{split}
    U(t_1,t_0)\, U(t_0\!+\!T,t_0)
    &= U(t_1\!+\!T,t_0\!+\!T)\,U(t_0\!+\!T,t_0) \\
    &= U(t_1\!+\!T,t_0) = U(t_1\!+\!T,t_1)\,U(t_1,t_0)\,,
\end{split}
\end{align}
which, by multipliying with $U(t_1,t_0)^{-1}$ from the left, implies
\begin{align}
    U(t_1\!+\!T,t_1) = U(t_1,t_0)\, U(t_0\!+\!T,t_0)\, U(t_1,t_0)^{-1}
\end{align}
With that, one obtains the transformation of the corresponding Floquet-Hamiltonian
\begin{equation}
    \widetilde H(t_1) = U(t_1,t_0)\, \widetilde H(t_0)\, U(t_1,t_0)^{-1}\,.
\end{equation}
Thus $\widetilde H(t_0)$ and $\widetilde H(t_1)$ are related by a similarity transformation. They share the same spectrum and therefore give the same stability information, while their eigenvectors and micromotion operators differ. The difference in the micromotion operators only reflects that the periodic part of the motion is measured from a different phase of the drive. This gauge freedom reflects that Floquet theory allows different partitions of the same dynamics into an effective exponential evolution and a periodic micromotion, without changing the physical motion. For further discussion see Refs.~\citenum{higashikawa2018floquetengineeringclassicalsystems, Bukov2015, Yakubovich1975}.

\section{Dyson Series for Commuting Matrices}\label{app:dyson commute}
If the Hamiltonian commutes with itself at different times, i.e., $[H(t), H(t')] = 0$ for all $t, t'$, the time-ordering becomes trivial, and the Dyson series reduces to the matrix exponential
\begin{equation}
    U(t) = \exp\left( \int_0^t H(t')\, \mathrm{d}t' \right)\,.
\end{equation}
This follows by noting that the $k$th term of the Dyson series becomes
\begin{equation}
    U^{(k)}(t) = \frac{1}{k!} \left( \int_0^t H(t')\, \mathrm{d}t' \right)^k,
\end{equation}
as all integrals are over symmetric domains in which permutations of the integrand commute.

However, this simplification does not apply to our problem. The system matrix $H(t)$ does not commute at different times
\begin{align}
    [H(t), H(t')] = (\alpha_\mp(t') - \alpha_\mp(t))
    \begin{bmatrix}
        1 & 0 \\ 0 & -1
    \end{bmatrix} \neq 0.
\end{align}

\section{Matching the Magnus and Dyson Expansions}\label{app:magnus-matching}
Starting from the relation between the Magnus and Dyson expansions used in the main text, $\exp\left(\sum_{k=1}^{\infty}\Omega^{(k)}(t)\right) = \sum_{k=0}^{\infty} U^{(k)}(t)$, we expand the exponential and match terms of equal order in $H$. Superscripts in parentheses denote the order in $H$ and should not be confused with powers of $\Omega$. Expanding the exponential to third order gives
\begin{align}\label{eq:magnus-exp-third}
\begin{split}
    \exp(\Omega(t))
    &= I + \Omega^{(1)}(t)
    + \left[\Omega^{(2)}(t) + \frac{1}{2}\left(\Omega^{(1)}(t)\right)^2\right] \\
    &\;\; + \left[\Omega^{(3)}(t) + \frac{1}{2}\left(\Omega^{(1)}(t)\Omega^{(2)}(t) + \Omega^{(2)}(t)\Omega^{(1)}(t)\right) + \frac{1}{6}\left(\Omega^{(1)}(t)\right)^3\right]
    + \mathcal{O}(\norm{H}^4)\,.
\end{split}
\end{align}
The Dyson series to the same order reads
\begin{equation}\label{eq:dyson-third}
    U(t) = I + U^{(1)}(t) + U^{(2)}(t) + U^{(3)}(t) + \mathcal{O}(\norm{H}^4).
\end{equation}
Matching terms of equal order in $H$ thus gives
\begin{align}\label{eq:magnus-matching}
\begin{split}
    \Omega^{(1)}(t) = U^{(1)}(t)\,, \\
    \Omega^{(2)}(t) + \frac{1}{2}\left(\Omega^{(1)}(t)\right)^2 = U^{(2)}(t)\,, \\
    \Omega^{(3)}(t) + \frac{1}{2}\left(\Omega^{(1)}(t)\Omega^{(2)}(t) + \Omega^{(2)}(t)\Omega^{(1)}(t)\right) + \frac{1}{6}\left(\Omega^{(1)}(t)\right)^3 = U^{(3)}(t)\,.
\end{split}
\end{align}
Solving these equations for the individual Magnus terms yields the third-order expressions used in the main text.

\section{Detailed Calculation of the Floquet Hamiltonian for Kapitza's Pendulum}\label{app:details}
To compute $\widetilde{H}$ we compute the individual terms. Starting with $\Omega^{(1)}$ We have
\begin{align}
    \Omega^{(1)}
    = \int_0^T
    \begin{bmatrix}
        0 & 1 \\
        \alpha_\mp(t') & 0
    \end{bmatrix}\;\d t'
        = T\begin{bmatrix}
    0 & 1 \\
    \mp \omega_0^2 & 0
\end{bmatrix}\,,
\end{align}
where we used
\begin{equation*}
    \mp\int_0^T \ddot{y}_\mathrm{P}(t') \;\d t' = 0\,,
\end{equation*}
as we integrate over one period of oscillation. Next we compute $\Omega^{(2)}_\mp$ via
\begin{align*}
    \Omega^{(2)} &= U^{(2)} - (\Omega^{(1)})^2 / 2\\
    &= \int_0^T\d t_1 \int_0^{t_1}\d t_2
    \begin{bmatrix}
        0 & 1 \\
        \alpha_\mp(t_1) & 0
    \end{bmatrix}
    \begin{bmatrix}
        0 & 1 \\
        \alpha_\mp(t_2) & 0
    \end{bmatrix}
    - \frac{T^2}{2}
    \begin{bmatrix}
        \mp \omega_0^2 & 0 \\ 0 & \mp\omega_0^2
    \end{bmatrix} \\
    &= \int_0^T\d t_1 \int_0^{t_1}\d t_2
    \begin{bmatrix}
        \alpha_\mp(t_2) & 0 \\
        0 & \alpha_\mp(t_1)
    \end{bmatrix}
    \pm \frac{T^2\omega_0^2}{2}
    \begin{bmatrix}
    1 & 0 \\ 0 & 1
    \end{bmatrix}\,.
\end{align*}
And with
\begin{align*}
    \int_0^T\d t_1 \int_0^{t_1}\d t_2 \; \alpha_\mp(t_2)
    &=  \int_0^T\d t_1 \int_0^{t_1}\d t_2 \; \left( \pm\frac{A\omega}{\ell} \cos(\omega t_2) \mp \omega_0^2 \right) \\
    &= \pm\int_0^T\d t_1\; \left( \frac{A}{\ell} \sin(\omega t_1) - t_1 \omega_0^2 \right)
    = \mp \frac{T^2\omega_0^2}{2}\,,
\end{align*}
where the first integral vanishes, and
\begin{align*}
    \int_0^T\d t_1 \int_0^{t_1}\d t_2 \; \alpha_\mp(t_1)
    & = \int_0^T\d t_1 \; t_1 \alpha_\mp(t_1) \\
    &=  \pm\int_0^T\d t_1\; \left( \frac{A\omega}{\ell} t_1\cos(\omega t_1) - t_1\omega_0^2 \right)
    = \mp \frac{T^2\omega_0^2}{2}\,,
\end{align*}
where again the first integral vanishes, we get
\begin{equation}
    \Omega^{(2)} = 0\,.
\end{equation}
So we only have $U^{(3)}(T)$ left to calculate, which is given by
\begin{align*}
    U^{(3)}(T)
    &= \int_0^T\d t_1 \int_0^{t_1}\d t_2 \int_0^{t_2}\d t_3
    \begin{bmatrix}
        0 & 1 \\ \alpha_\pm(t_1) & 0
    \end{bmatrix}
    \begin{bmatrix}
        0 & 1 \\ \alpha_\pm(t_2) & 0
    \end{bmatrix}
    \begin{bmatrix}
        0 & 1 \\ \alpha_\pm(t_3) & 0
    \end{bmatrix} \\
    &= \int_0^T\d t_1 \int_0^{t_1}\d t_2 \int_0^{t_2}\d t_3
    \begin{bmatrix}
        0 & \alpha_\pm(t_2) \\ \alpha_\pm(t_3) \alpha_\pm(t_1) & 0
    \end{bmatrix}\,.
\end{align*}
And we have
\begin{align*}
\begin{split}
    \int_0^T\d t_1 \int_0^{t_1}\d t_2 \int_0^{t_2}\d t_3\; \alpha_\mp(t_2)
    &= \int_0^T\d t_1 \int_0^{t_1}\d t_2 \; t_2 \alpha_\pm(t_2) \\
    &= \pm\int_0^T\d t_1 \int_0^{t_1}\d t_2 \; t_2 \left( \frac{A\omega^2}{\ell} \cos(\omega t_2) - \omega_0^2\right) \\
    &= \pm \frac{A\omega^2}{\ell} \int_0^T \d t_1 \underbrace{ \int_0^{t_1} \d t_2\; t_2\cos(\omega t_2)}_{\frac{t_1\sin(\omega t_1)}{\omega} + \frac{\cos(\omega t_1)}{\omega^2} -\frac{1}{\omega^2} } \mp \frac{\omega_0^2 T^3}{6} \\
    &=  \mp \frac{\omega_0^2 T^3}{6} \mp \frac{2A T}{\ell}\,,
\end{split}
\end{align*}
where we used that $\int_0^T\d t_1 \frac{t_1 \sin(\omega t_1)}{\omega} = -\frac{T}{\omega^2}$.
For the other term we get
\begin{align*}
\begin{split}
    \int_0^T&\d t_1 \int_0^{t_1}\d t_2 \int_0^{t_2}\d t_3\;\alpha_\mp(t_1) \alpha_\mp(t_3) \\
    &= \int_0^T\d t_1 \; \left( \pm \frac{A\omega^2}{\ell}\cos(\omega t_1) \mp \omega_0^2 \right) \overbrace{\int_0^{t_1} \d t_2 \int_0^{t_2}\d t_3 \left( \pm \frac{A\omega^2}{\ell}\cos(\omega t_3) \mp \omega_0^2 \right)}^{\mp \frac{A}{\ell} \cos(\omega t_1) \pm \frac{A}{\ell} \mp \frac{\omega_0^2 t_1^2}{2}} \\
    &= - \frac{A^2\omega^2 T}{2\ell^2} - \frac{2A\omega_0^2 T}{\ell} + \frac{\omega_0^4 T^3}{6}\,.
\end{split}
\end{align*}
With that, we calculate
\begin{align}
\begin{split}
    \Omega^{(3)}
    &= U^{(3)} - \frac{1}{6}(\Omega^{(1)})^3 \\
    &=
    \begin{bmatrix}
        0 & \mp \frac{\omega_0^2 T^3}{6} \mp \frac{2AT}{\ell} \\
        - \frac{A^2\omega^2 T}{2\ell^2} - \frac{2A\omega_0^2 T}{\ell} + \frac{\omega_0^4 T^3}{6} & 0
    \end{bmatrix}
    - \frac{1}{6}
    \begin{bmatrix}
        0 & T \\ \mp \omega_0^2 T & 0
    \end{bmatrix}^3 \\
    &=
    \begin{bmatrix}
        0 & \mp \frac{\omega_0^2 T^3}{6} \mp \frac{2AT}{\ell} \\
        - \frac{A^2\omega^2 T}{2\ell^2} - \frac{2A\omega_0^2 T}{\ell} + \frac{\omega_0^4 T^3}{6} & 0
    \end{bmatrix}
    - \frac{1}{6}
    \begin{bmatrix}
        0 & \mp \omega_0^2 T^3 \\ \omega_0^4 T^3 & 0
    \end{bmatrix} \\
    &=
    \begin{bmatrix}
        0 & \mp \frac{2AT}{\ell} \\
        - \frac{A^2\omega^2 T}{2\ell^2} - \frac{2A\omega_0^2 T}{\ell} & 0
    \end{bmatrix}\,.
\end{split}
\end{align}
In total, we get
\begin{align}
\begin{split}
    \widetilde{H} &\approx \frac{\Omega^{(1)} + \Omega^{(2)} + \Omega^{(3)}}{T}
    =
    \begin{bmatrix}
        0 & 1 \mp \frac{2A}{\ell} \\
        \mp \omega_0^2 - \frac{A^2\omega^2}{2\ell^2} - \frac{2A\omega_0^2}{\ell} & 0
    \end{bmatrix}\,,
\end{split}
\end{align}
which can be further approximated for small amplitude $2A/\ell \ll 1$ as
\begin{align}
    \widetilde{H} \approx
    \begin{bmatrix}
        0 & 1 \\
        - \frac{1}{2}\left(\frac{A\omega}{\ell}\right)^2 \mp \omega_0^2 & 0
    \end{bmatrix}\,.
\end{align}

\section*{Acknowledgments}
We acknowledge funding by the Austrian Science Fund (FWF) [10.55776/P36903]. The idea for this problem was developed during the Austrian preliminaries of the PLANCKS competition, \href{https://plancks.at}{PLANCKS Austria}. We thank \href{https://pauli-physics.at}{pauli-physics} for organizing the event and for the opportunity to contribute a problem. We are also grateful to Prof.~Enrico Arrigoni for stimulating comments and a careful review of the manuscript.

% Add references if needed
%\bibliographystyle{apsrev4-2}
\bibliography{main}

\end{document}